\documentclass{webofc}
\usepackage[varg]{txfonts}   
\begin{document}
\title{Exclusive Diffraction at the LHC \\
On behalf of the ATLAS, LHCb, CMS and TOTEM experiments}
%
%

\author{\firstname{Christophe} \lastname{Royon}\inst{1,3}\fnsep\thanks{\email{christophe.royon@ku.edu}} }

\institute{Department of Physics and Astronomy, The University of Kansas, Lawrence, USA 
          }

\abstract{%
 In this report, we describe the most recent results on exclusive diffraction from the ATLAS, CMS, LHCb, TOTEM experiments at the LHC concerning
 exclusive pions, $J/\Psi$, $\Psi(2S)$, dilepton, diphoton, $WW$ productions and prospects concerning the search for anomalous couplings and axion-like particle
 production.
}
\maketitle
\section{Introduction: Definition of exclusive diffraction}
\label{intro}

\begin{figure}[h]
\centering
\includegraphics[width=1.in,clip]{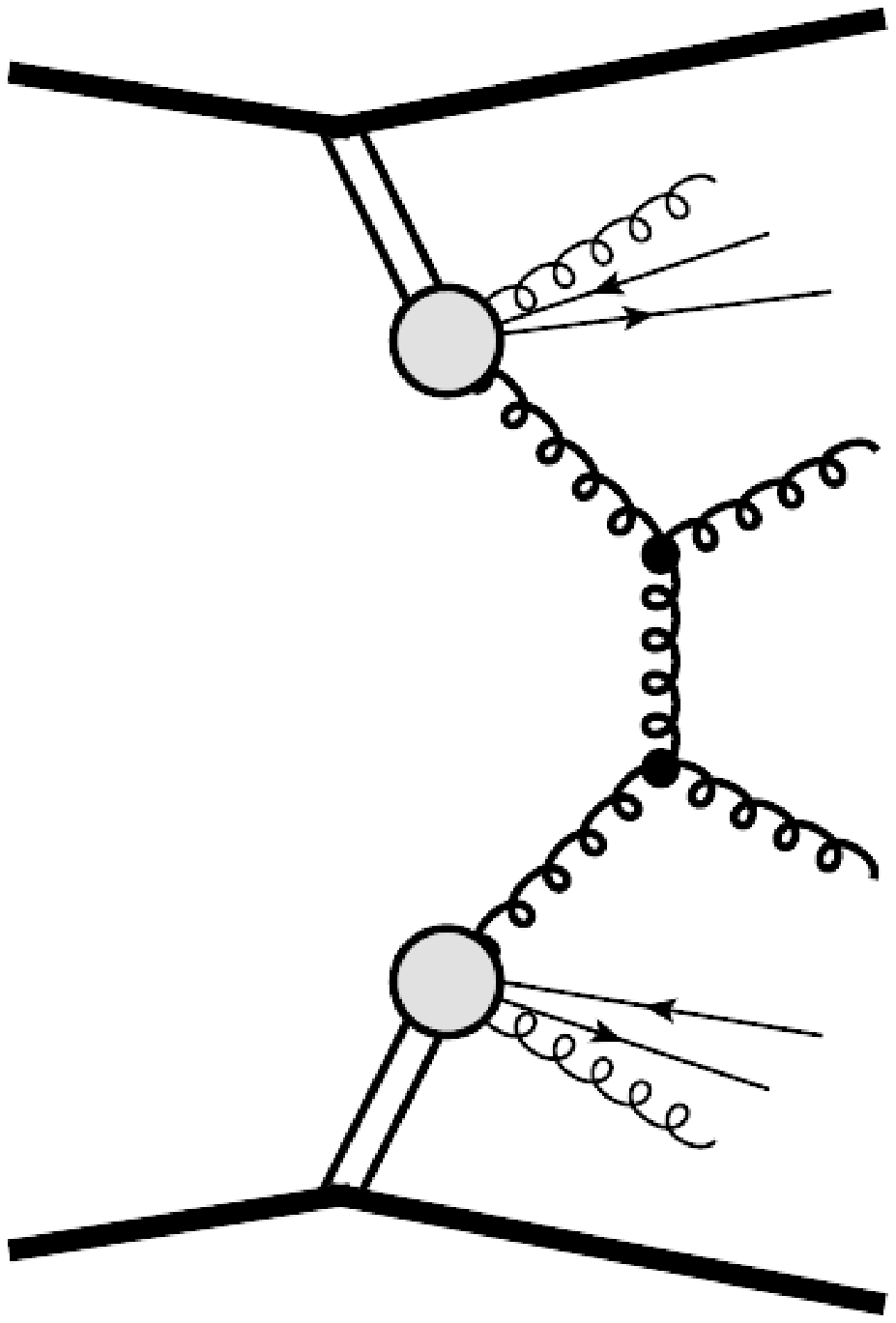}
\includegraphics[width=1.in,clip]{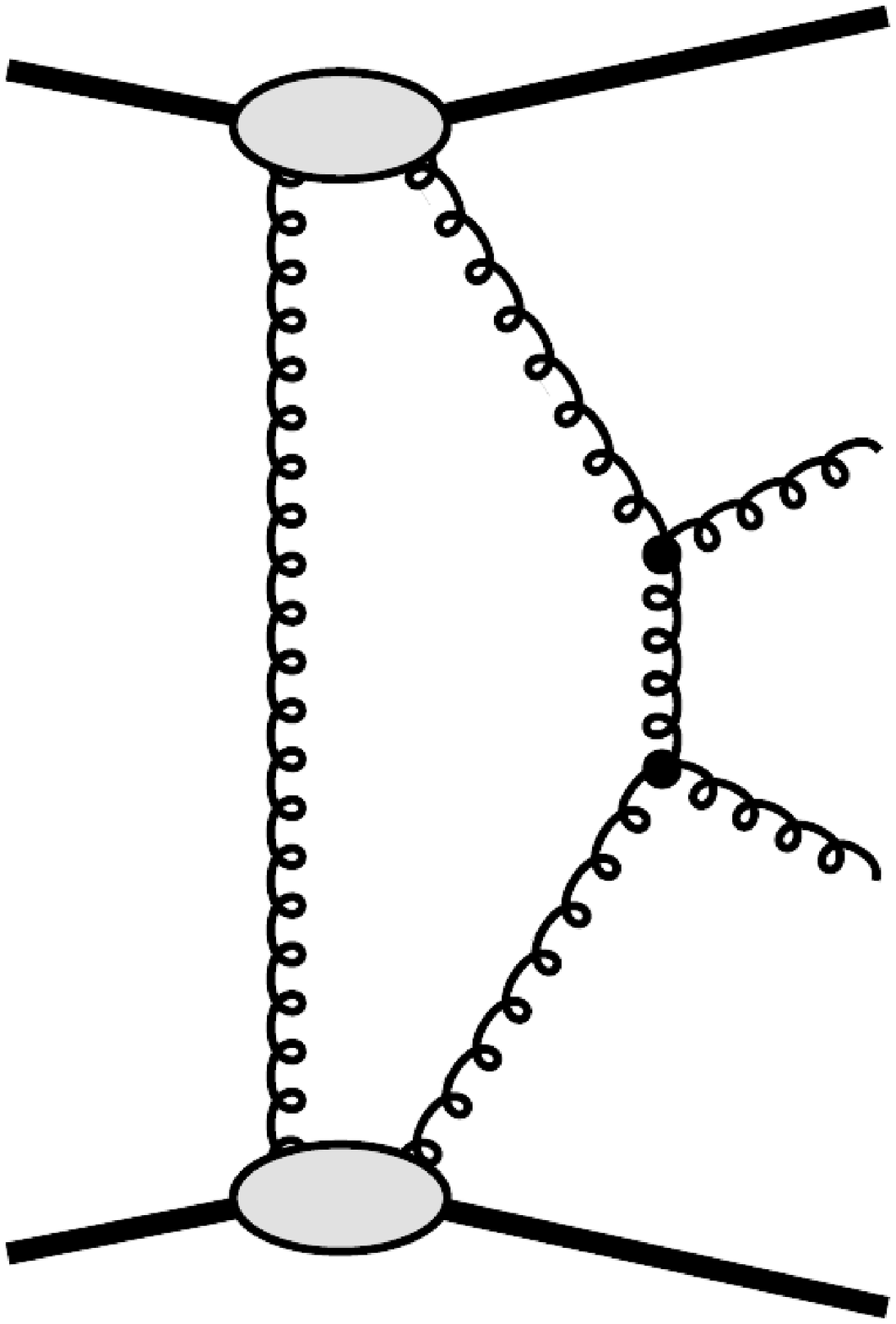}
\includegraphics[width=1.4in,clip]{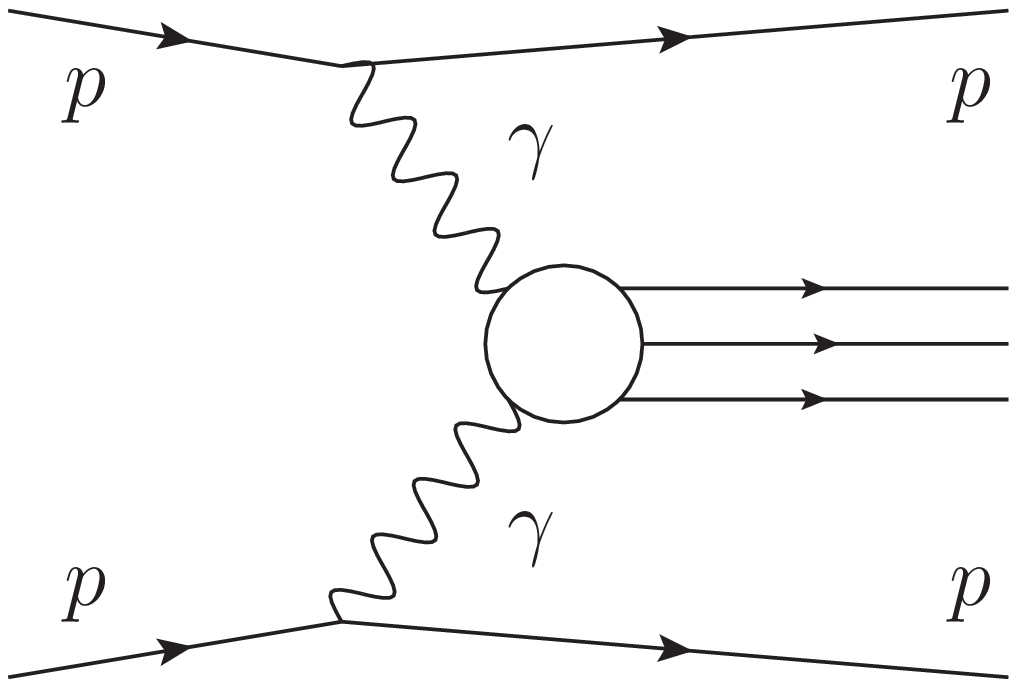}
\includegraphics[width=1.3in,clip]{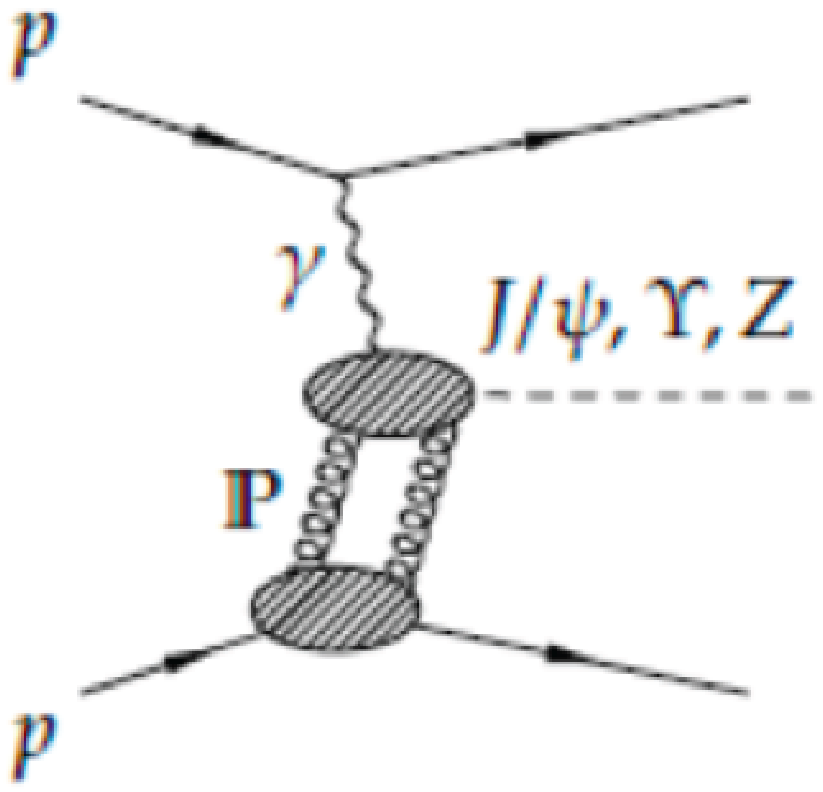}
\caption{Diagrams of inclusive and exclusive diffractive processes.}
\label{fig1}       
\end{figure}

We will start this report by defining what we call ``Exclusive" diffraction. The first left diagram of Fig.~\ref{fig1} corresponds
to Double Pomeron Exchange in inclusive diffraction. In this event, both protons are intact in the final state and two Pomerons
are exchanged. Gluons are extracted from each Pomeron in order to produce jets (or diphotons, $W$s...). Some energy is ``lost" in Pomeron remnants. The three 
other diagrams in Fig.~\ref{fig1} are exclusive in the sense that the full energy is used to produce the diffractive object. In other
word, there is no energy loss in Pomeron remnants. The second diagram corresponds to exclusive diffraction~\cite{KMR}, the third one
to photon exchanges and the last one to photon Pomeron exchanges that produce vector mesons. In the following, we will discuss the
measurements at the LHC corresponding to these different diagrams.

\section{Low mass exclusive diffraction}

In this section, we will focus on low mass exclusive diffraction results from LHCb and CMS, namely on vector meson and pion production.

\subsection{Measurement of central exclusive production in LHCb}

\begin{figure}[h]
\centering
\includegraphics[width=4.5in,clip]{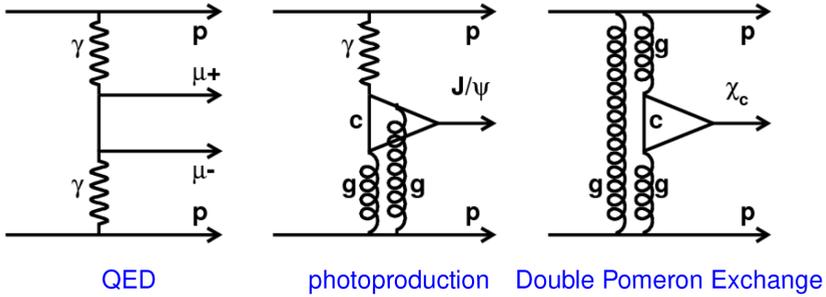}
\caption{Central exclusive production in LHCb.}
\label{fig2}       
\end{figure}

\begin{figure}[h]
\centering
\includegraphics[width=2.5in,clip]{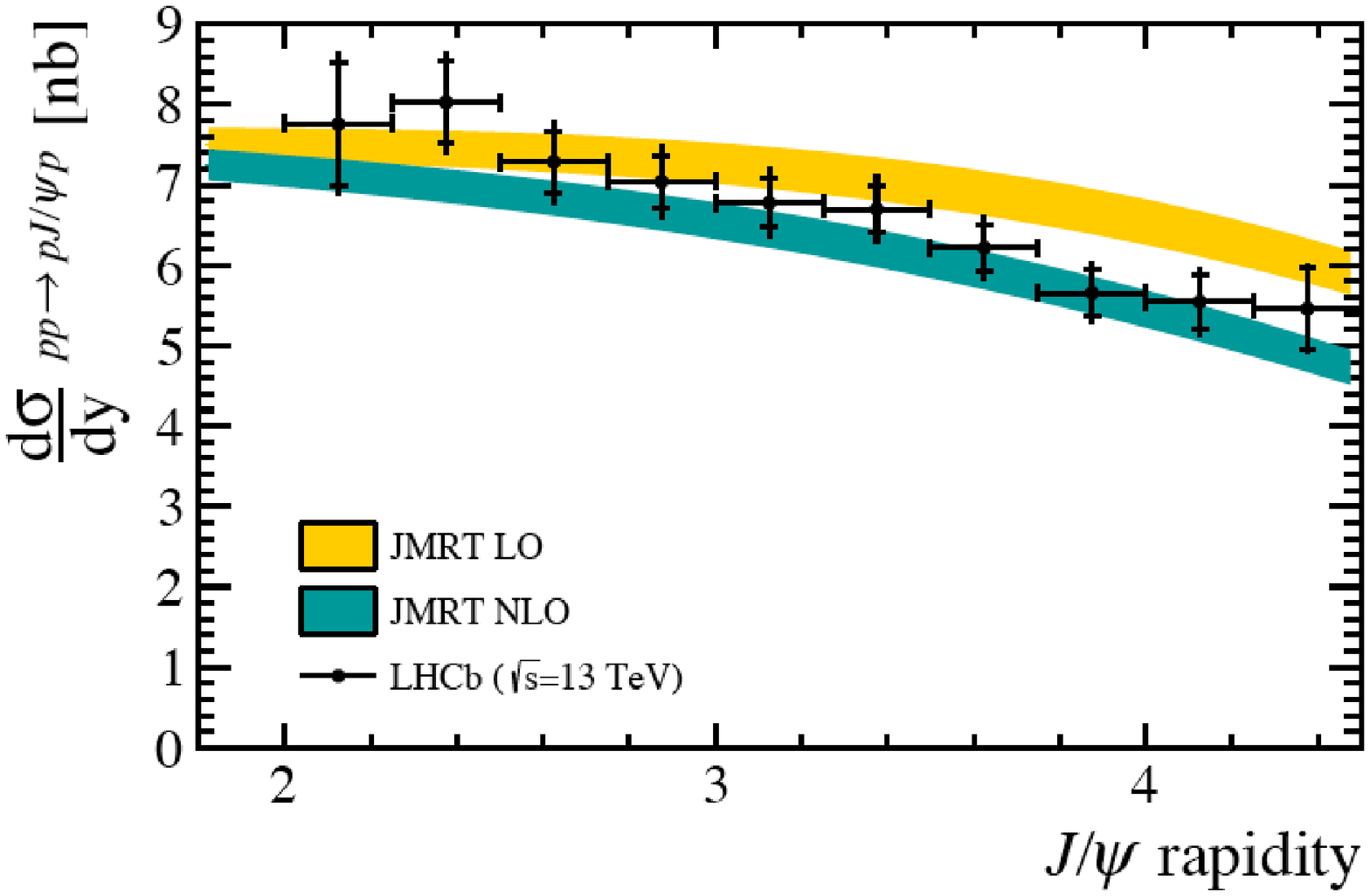}
\includegraphics[width=2.5in,clip]{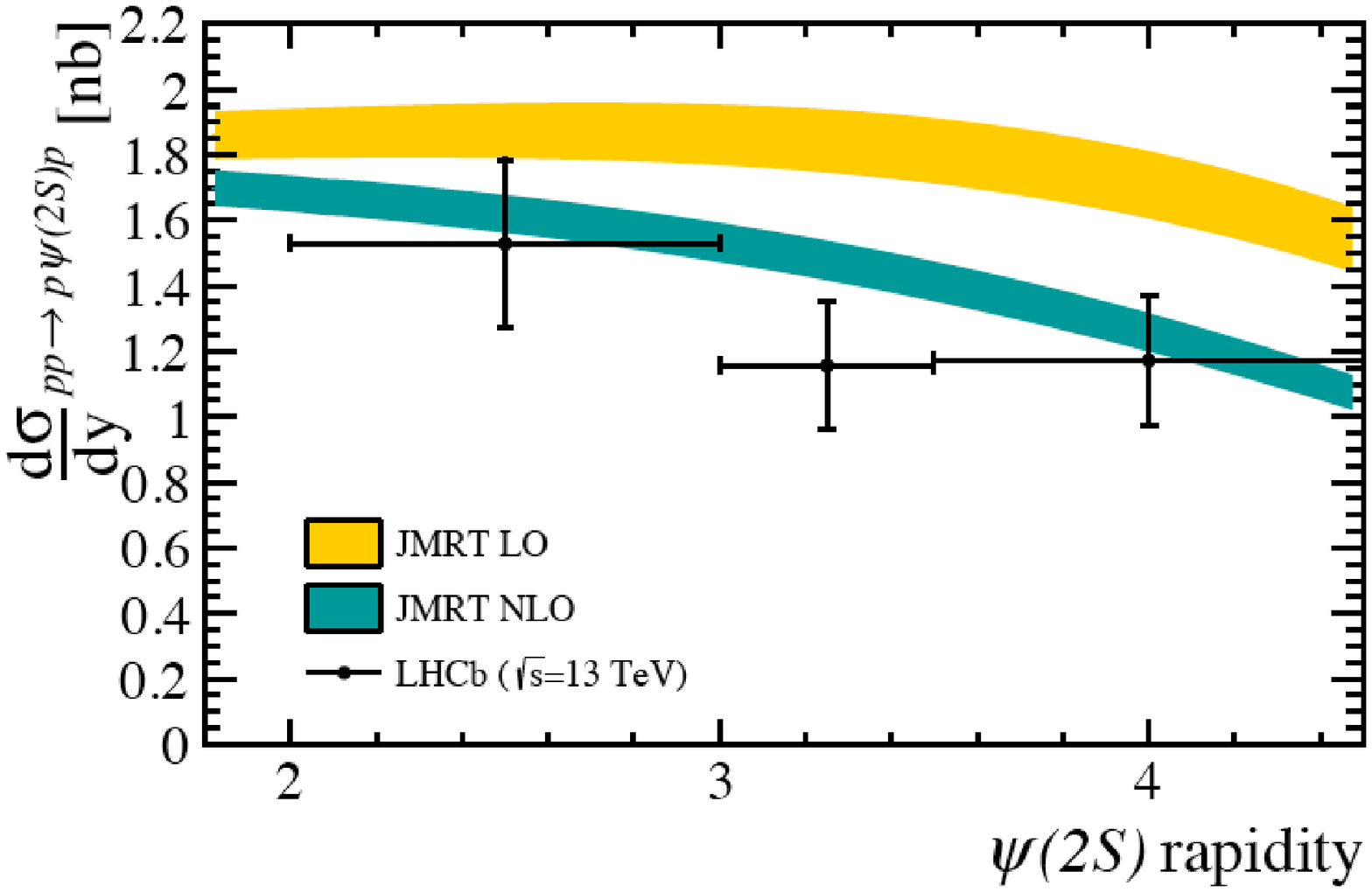}
\caption{Measurement of the $J/\Psi$ and $\Psi(2S)$  cross sections as a function of rapidity using  13 TeV data including the
new HERSCHEL detector.}
\label{fig3}       
\end{figure}

\begin{figure}[h]
\centering
\includegraphics[width=2.5in,clip]{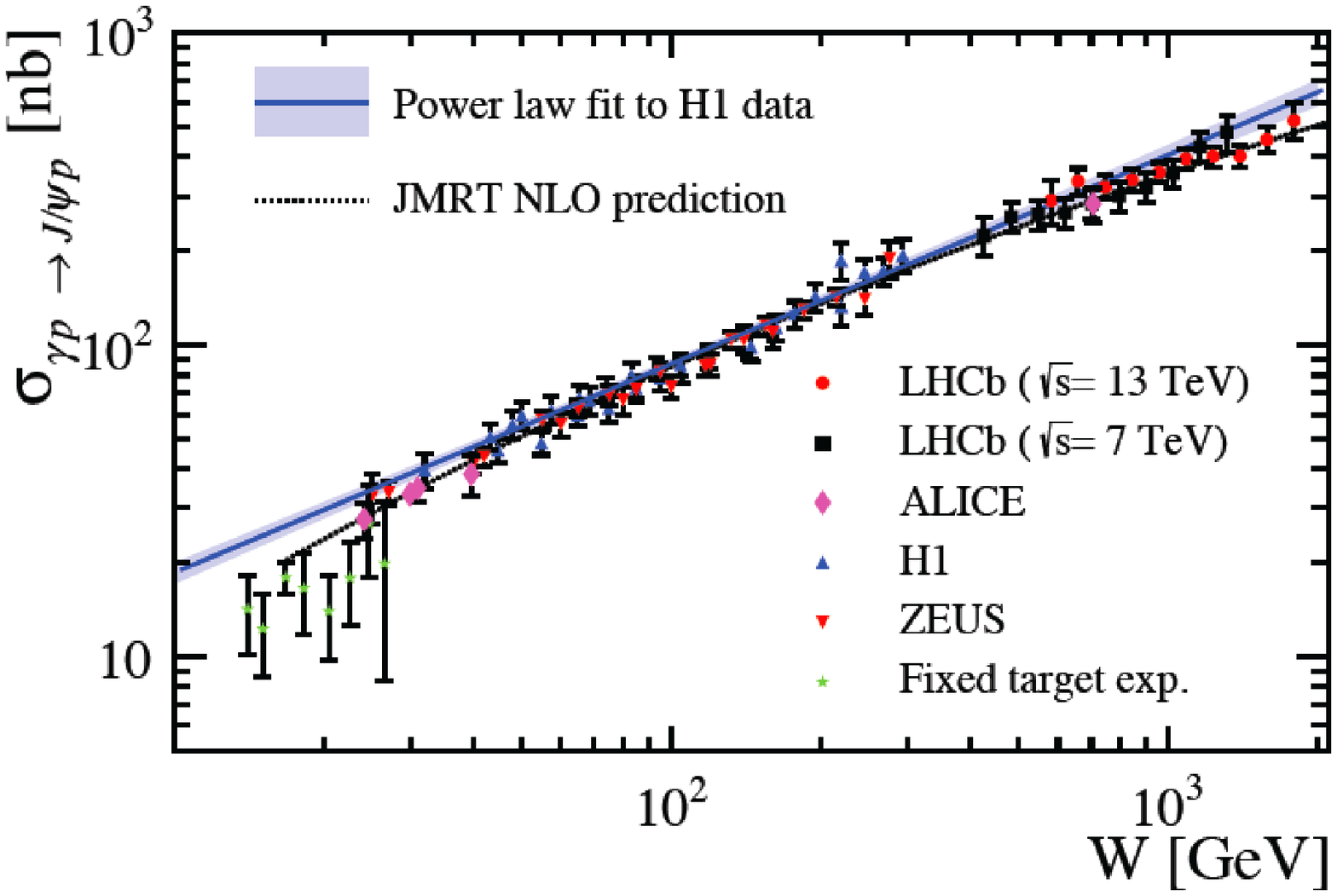}
\includegraphics[width=2.5in,clip]{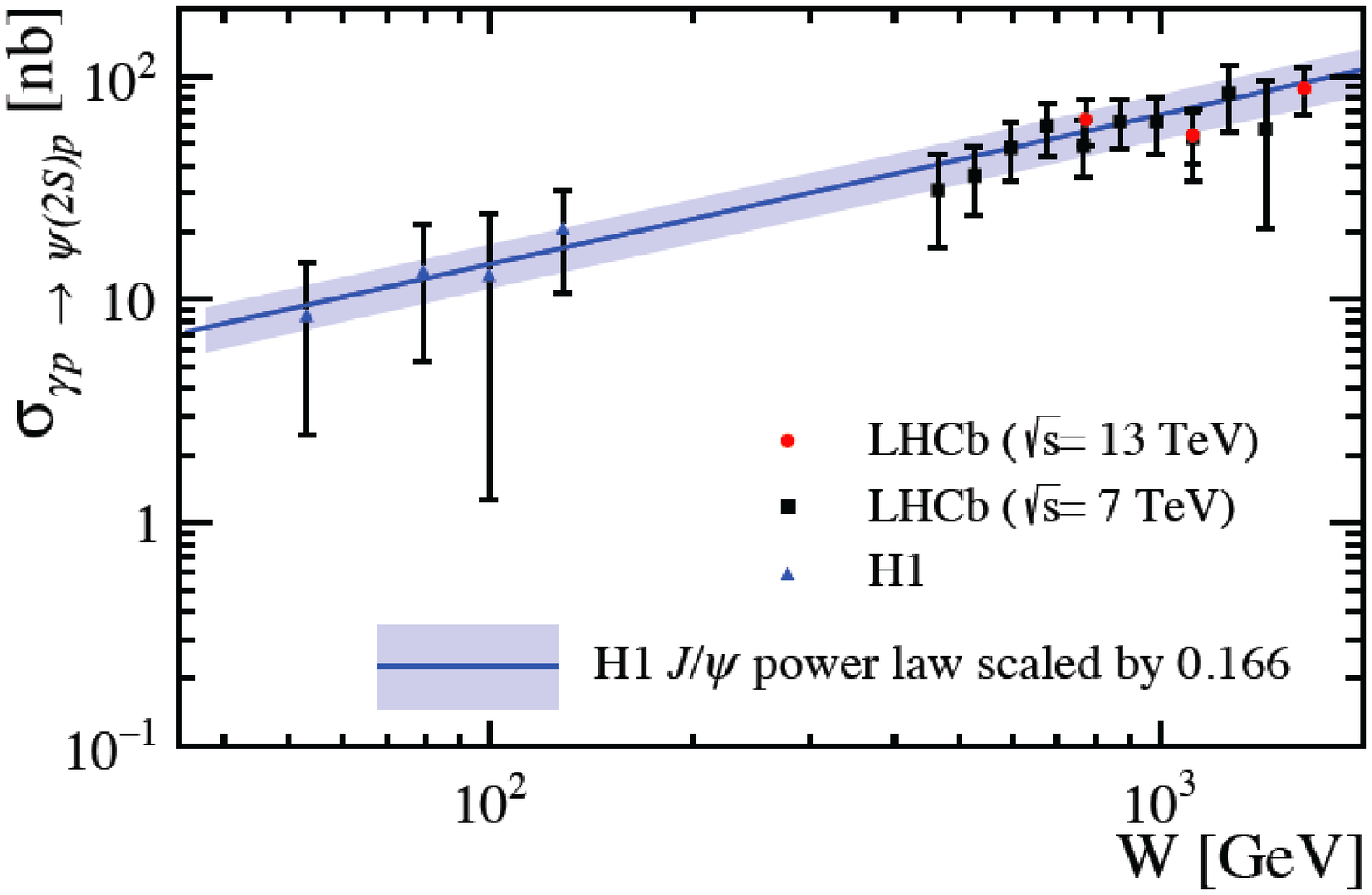}
\caption{Measurement of the $\gamma p \rightarrow J/\Psi p $ and $\gamma p \rightarrow \Psi(2S) p$  cross sections as a function of energy using 7 and 13 TeV data
compared to the JMRT NLO parametrization. Data cannot be described by a simple power law.}
\label{fig4}       
\end{figure}

In Fig.~\ref{fig2}, we display three diagrams that correspond to exclusive productions in LHCb. The first left diagram corresponds to the QED production of di-muon, the middle one the photoproduction of $J/\Psi$, and the right one the inclusive Double Pomeron Exchange production of $\chi_C$ mesons. The exclusive production of $J/\Psi$ is strongly sensitive to the gluon distribution in the proton since it appears as a square dependence in the amplitude. The signal is to measure the $J/\Psi$ in the central part of LHCb and the background is due to inclusive diffractive processes where the Pomeron remnants are not detected since exclusive processes are selected via the rapidity gap method. The issue is the gap detection since, for some events, the Pomeron remnants are outside detector acceptance. 

In order to improve the coverage at high rapidities, the LHCb collaboration installed a new detector called HERSCHEL (High
Rapidity Shower Counter for LHCb).  HERSCHEL is installed on both sides of LHCb and allows a coverage in pseudo-rapidity of $-10.0<\eta <-3.5$ and $5.0 < \eta < 10.0$ which leads to a coverage of LHCb of  $-10.0<\eta <-1.5$ and $2.0 < \eta < 10.0$. HERSCHEL was used to select exclusive data
at 13 TeV, and about 14753 $J/\Psi$ and 440 $\Psi(2S)$ were selected in 204 pb$^{-1}$ of data with a respective purity of 75.5$\pm$1.5\% and 72.6$\pm$6.1\%.
The measured cross sections are $\sigma (J/\Psi \rightarrow \mu^+ \mu^-) (2 <\eta < 4.5)=435 \pm 18 \pm 11 \pm 17$ pb, and
$\sigma (\Psi(2S)\rightarrow \mu^+ \mu^-) (2 <\eta < 4.5)=11.1 \pm 1.1 \pm 0.3 \pm 0.4$ pb.
The differential cross sections for $J/\Psi$ and $\Psi(2S)$ compared between 7 and 13 TeV data are shown in Fig.~\ref{fig3}~\cite{lhcbjpsi}. The results are in agreement with NLO calculations and the best PDF parametrisation is from JMRT at NLO~\cite{jmrt}.

This measurement allows LHCb to compare with previous measurements performed by the H1 and ZEUS collaborations at HERA using the following simple relation between HERA and LHCb measurements
\begin{eqnarray}
\frac{d \sigma}{d y_{pp \rightarrow pVp}} = r(y) \left[ k_+ \frac{dn}{dk_+} \sigma_{\gamma p \rightarrow Vp}(W_+) +
k_- \frac{dn}{dk_-} \sigma_{\gamma p \rightarrow Vp}(W_-)  \right]
\end{eqnarray}
where the $pp \rightarrow pVp$ is measured by LHCb, $\sigma(W_-)$ (resp. $\sigma(W_+)$) at HERA, and $\sigma(W_+)$ (resp. $\sigma(W_-)$) is extracted from the previous formula. The results are shown in Fig.~\ref{fig4}~\cite{lhcbjpsi} and a simple power law does not lead to a good description of data.

\subsection{Measurement of exclusive pion production in CMS}

\begin{figure}[h]
\centering
\includegraphics[width=2.in,clip]{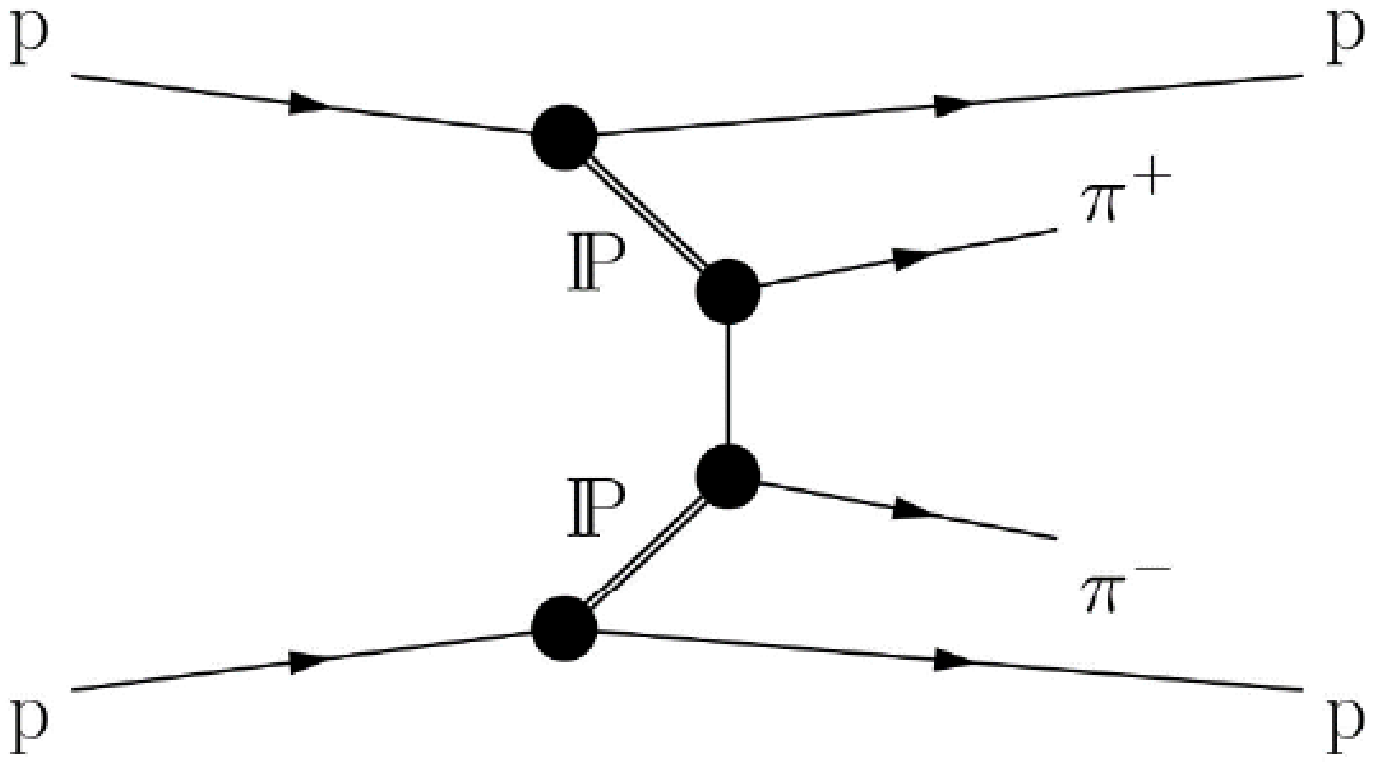}
\includegraphics[width=2.in,clip]{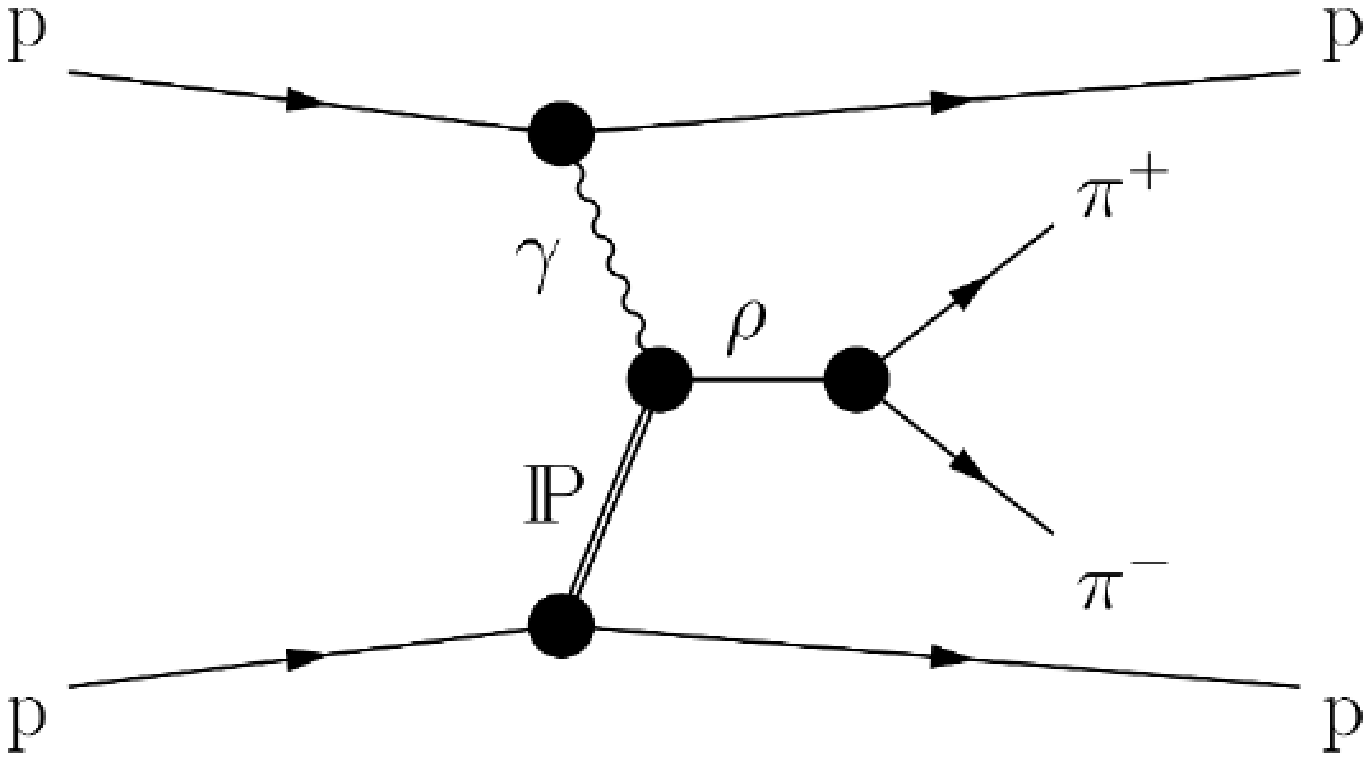}
\caption{Leading diagrams leading to exclusive pion production.}
\label{fig5}       
\end{figure}

\begin{figure}[h]
\centering
\includegraphics[width=4.5in,clip]{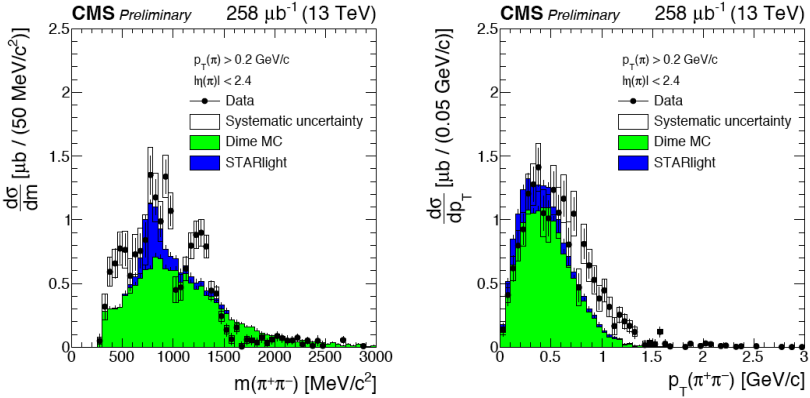}
\caption{Measurement of exclusive pion cross section by the CMS collaboration as a function of the di-pion mass in linear and logarithmic scales.}
\label{fig6}       
\end{figure}

In this section, we will describe the measurement of exclusive pion production by the CMS collaboration~\cite{cmspion}.  The two diagrams that are dominant at low masses are shown in Fig.~\ref{fig5}. The left diagram corresponds to double Pomeron exchange and is implemented in the DIME Monte carlo~\cite{dime} whereas the right diagram corresponds to $\gamma$-Pomeron exchanges producing a $\rho$ meson that decays into a pion pair as implemented in the STARLIGHT Monte Carlo~\cite{starlight}. The measurement was performed using low pile up runs since the production cross section is high and not much accumulated luminosity is needed. The same measurement in a high pile up environment would be challenging because of the low $p_T$ track requirement. The experimental signature is to request two opposite tracks from the same primary vertex and no additional signal in the calorimeter outside the two pions with $p_T(\pi)>0.2$ GeV and $|y(\pi)|<2$. The background is computed using same sign events directly in data.

The measurement of the di-pion cross section is shown in Fig.~\ref{fig6} as a function of $m_{\pi \pi}$ and $p_T(\pi \pi)$~\cite{cmspion}. We see a discrepancy between the measurements and the sum of the DIME and STARLIGHT Monte Carlo expectations both in shape and normalization. Both Monte Carlos do not contain proton dissociation and it might be a reason for most of the discrepancy (since the measurements require the existence of a rapidity gap in the CMS detector, it is impossible to distinguish experimentally between the cases when the protons are intact or dissociate). The exclusive pion cross section was measured to be $\sigma_{\pi^+ \pi^-} (13 TeV) = 19.0 \pm 0.6 (stat) \pm 3.2 (syst) \pm 0.01 (lumi)$ $\mu$b;
$\sigma_{\pi^+ \pi^-} (5.02 TeV) = 19.6 \pm 0.4 (stat) \pm 3.3 (syst) \pm 0.01 (lumi)$ $\mu$b.

\section{Exclusive measurements at high mass and photon exchange processes}

In this section, we will describe the results on exclusive production at high mass together with the ones on photon exclusive exchange processes.

\subsection{ATLAS and CMS results on exclusive $WW$ production}
ATLAS and CMS measured the exclusive production of a pair of $W$ bosons $pp \rightarrow pWWp$ using the rapidity gap technique. This measurement  is motivated by the search for quartic anomalous $\gamma \gamma WW$ couplings defined as additional terms in the SM lagrangian
\begin{eqnarray}
\mathcal{L}^0_6 &\sim& \frac{-e^2}{8} \frac{a_0^W}{\Lambda^2} F_{\mu \nu}
F^{\mu \nu} W^{+ \alpha} W^{-}_{\alpha} - \frac{e^2}{16 \cos^2(\theta_W)}
\frac{a_0^Z}{\Lambda^2}  F_{\mu \nu} F^{\mu \nu} Z^{\alpha} Z_{\alpha} 
\nonumber \\
\mathcal{L}^C_6 &\sim& \frac{-e^2}{16} \frac{a_C^W}{\Lambda^2} F_{\mu
\alpha}
F^{\mu \beta} (W^{+ \alpha} W^{-}_{\beta} + W^{- \alpha} W^{+}_{\beta}) 
 - \frac{e^2}{16 \cos^2(\theta_W)}
\frac{a_C^Z}{\Lambda^2}  F_{\mu \alpha} F^{\mu \beta} Z^{\alpha} Z_{\beta}
\nonumber
\end{eqnarray}
All parameters are equal to 0 in the SM. 

This search has to be performed using the highest possible integrated luminosity at the LHC since exclusive $WW$ events are rare which means at high pile up at the LHC. The CMS and ATLAS collaborations used respectively 19.7 fb$^{-1}$ and 20.2 fb$^{-1}$ at a center-of-mass of 8 TeV (and CMS an additional 5.05 fb$^{-1}$ at 7 TeV). The exclusivity selection requires opposite sign electron and muon originating from the common primary vertex and no extra track from that vertex. In addition, $M_{e \mu}>20$ GeV to avoid the low mass resonance region and $P_T^{e \mu}>30$ GeV to remove Drell-Yan and $\gamma \rightarrow \tau \tau$ events. 

CMS and ATLAS measured the exclusive $WW$ production and obtained  $\sigma (pp \rightarrow pWWp \rightarrow p \mu e p) =2.2 ^{+3.3}_{-2.0}$ fb at 7 TeV 
(SM: 4.0  $\pm$ 0.7 fb) ,
$\sigma (pp \rightarrow pWWp \rightarrow p \mu e p) =10.8 ^{+5.1}_{-4.1}$ fb at 
8 TeV (SM: 6.2
$\pm$ 0.5 fb) for CMS after correction for proton dissociation, and
$\sigma=6.9 \pm 2.2 (stat) \pm 1.4 (syst)$ fb)  for ATLAS (SM: 4.4
$\pm$ 0.3 fb).
The observed significance for 7 and 8 TeV combination: is 3.4 $\sigma$ for
CMS and  3.0 $\sigma$ for ATLAS~\cite{wwatlascms}

\begin{figure}[h]
\centering
\includegraphics[width=5.in,clip]{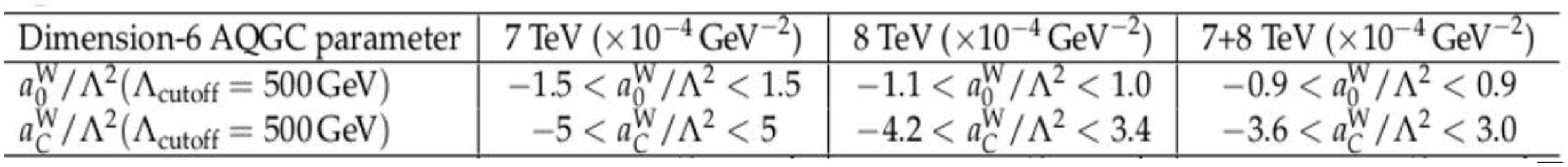}
\includegraphics[width=5.in,clip]{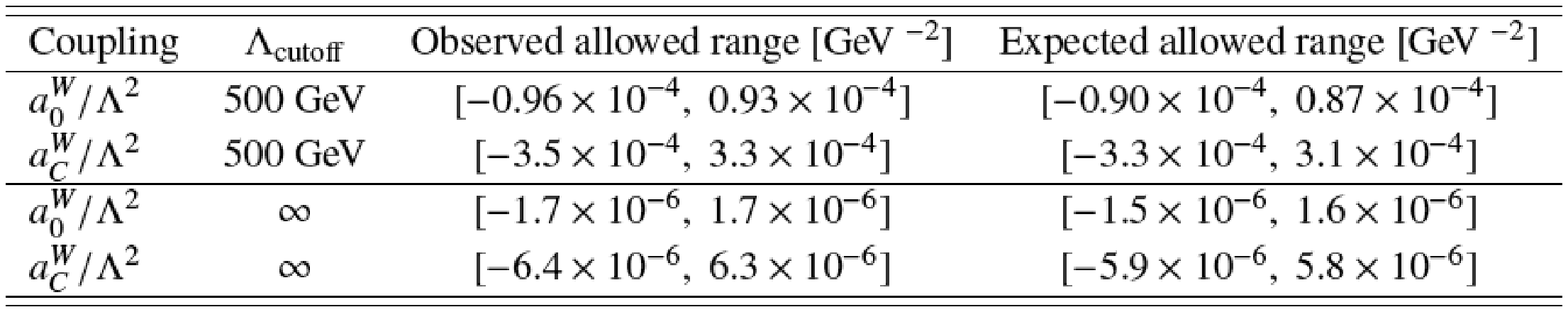}
\caption{Limits on $\gamma \gamma WW$ quartic anomalous couplings by the CMS and ATLAS collaborations..}
\label{fig7}       
\end{figure}

Since data are in agreement with the standard model, it is possible to obtain the most stringent limits to date on $\gamma \gamma WW$  quartic anomalous coupling and the results are given in Fig.~\ref{fig7}. It is worth noticing that these limits can potentially be improved by about two orders of magnitude measuring the intact protons in the final state using the CT-PPS and AFP detectors~\cite{oldaww}.

\subsection{Exclusive di-lepton production in ATLAS  and in CMS-TOTEM (PPS)}
The ATLAS Forward Proton~\cite{afp} (AFP) and CMS-TOTEM Precision Proton Spectrometer~\cite{ctpps} (PPS) detectors were installed recently in order to detect and measure intact protons in the final state that leads to a possible better identification of exclusive events. The LHC  magnets bend the scattered protons outside the beam envelope. The roman pots detectors are located at about 210-220 m from the center part of the CMS or ATLAS detectors and cover a region in diffractive mass between typically 350 and 2000 GeV depending on the beam lattice.
The PPS detector started taking data in 2016 and could accumulate about 15 fb$^{-1}$ in 2016, and about 115 fb$^{-1}$ between 2016 and 2018. While the full data set is still being analyzed, we will describe the first observation of exclusive di-leptons using 9.4 fb$^{-1}$.

\begin{figure}[h]
\centering
\includegraphics[width=5.in,clip]{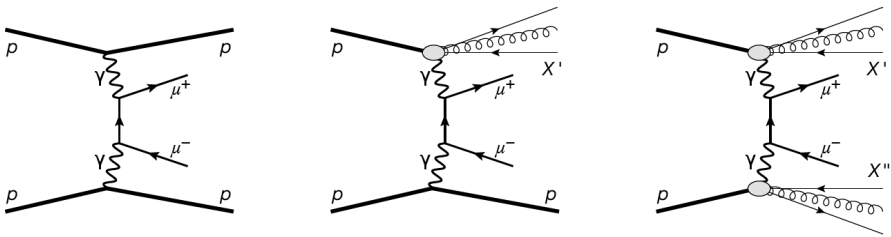}
\caption{Diagrams leading to di-muon production via photon exchanges..}
\label{fig8}       
\end{figure}

The ATLAS and CMS collaborations measured in 2016 the exclusive di-lepton production with two different strategies. The corresponding diagrams are shown in Fig.~\ref{fig8}. The LHC is turned into a $\gamma \gamma$ collider and the flux of quasi-real photons is computed using the Equivalent Photon Approximation (EPA). ATLAS did not have the possibility of detecting intact protons at the time of this measurement, and they selected exclusive $\mu \mu$ production using the rapidity gap technique. The exclusivity requirement in presence of pile up is that zero additional track exists within 1 mm of the $\mu^+ \mu^-$ vertex. From the rapidity gap technique, it is impossible to know if the protons are dissociate or not and the measurement corresponds to the three diagrams shown in Fig.~\ref{fig8}. The ATLAS collaboration fitted the $\mu \mu$ acoplanarity distribution at 13 TeV using 3.2 fb$^{-1}$ of data in order to extract the fiducial cross section for $p_T^{\mu} >$ 6 GeV ($12 < m_{\mu \mu}<$30 GeV) and for
$p_T^{\mu}>$ 10 GeV ($30 < m_{\mu \mu}<$70 GeV) and found
$\sigma_{\gamma \gamma \rightarrow \mu \mu}^{excl. fid} = 3.12 \pm 0.07 (stat)
\pm 0.10 (syst)$ pb~\cite{atlasmumu}. The comparison with the Superchic Monte Carlo~\cite{superchic} indicates an insufficient  suppression in Monte Carlo due to absorptive effects.

\begin{figure}[h]
\centering
\includegraphics[width=5.in,clip]{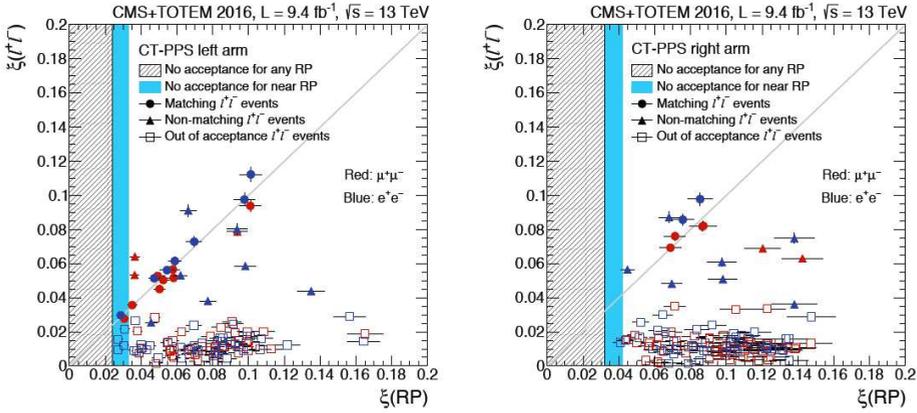}
\caption{Correlation between the $\xi$ values computed using the TOTEM roman pots and the di-lepton measured in CMS. The 20 semi-exclusive events are indicated in red. The left (right) plot displays the left (right) arm of TOTEM.}
\label{fig9}       
\end{figure}

The PPS collaboration measured exclusive di-lepton production by tagging one intact proton in the final state. This is the first time the semi-exclusive di-lepton processes are measured with proton tag. In Fig.~\ref{fig8}, the two left diagrams correspond to the signal whereas the rightmost diagram is part of the background. The reason that only one proton is requested to be tagged is that less than one event is expected for double tagged events with about 10 fb$^{-1}$ of data due to the mass acceptance above about 350 GeV for the forward proton detectors. A pair of opposite sign muons or electrons with $p_T>50$ GeV and $M_{l
l}>110$ GeV above the $Z$ boson peak is requested.
In order to suppress background, there is a veto on additional tracks around the di-lepton
vertex (within 0.5 mm) and leptons are required to be back-to-back, $|1 - \Delta
\Phi/\pi|<0.006$ for electrons (0.009 for muons).
The main background is due to Drell-Yan di-lepton production with the intact proton originating from pile up events. This background is estimated using Drell-Yan $Z$ events in data and extrapolating from the $Z$ peak region to our exclusive di-lepton signal region. 40 events (17 $\mu \mu$ and 23 $ee$) are found with protons in the PPS acceptance and 20 (12 $\mu \mu$ and 8 $ee$) show a less than $2 \sigma$ matching between the values of $\xi$ computed using the TOTEM roman pots and using the di-lepton measured in CMS as shown in Fig.~\ref{fig9}~\cite{ctppsnote}. This leads to a significance larger than 5$\sigma$ to observe  20 events for a
background of $3.85$ ($1.49\pm 0.07 (stat) \pm 0.53 (syst)$ for $\mu \mu$ and 
$2.36\pm 0.09 (stat) \pm 0.47 (syst)$ for $ee$). As expected, no event was double tagged with an intact proton on each side.


\subsection{Exclusive di-photon production via photon exchanges}
The ATLAS collaboration recently measured exclusive di-photon production in $Pb Pb$ production at the LHC. Exclusive di-photons can be produced via photon exchanges and the cross section is enhanced by a factor $Z^4$ with respect to the $pp$ case. In 480 $\mu$b$^{-1}$ of $PbPb$ data at $\sqrt{s}=$5.02 TeV in 2015, 13 events were observed for an estimated background of 2.6$\pm$0.7. Uisng 1.7 nb$^{-1}$ of data in 2018, 59 events were observed for a background of 12$\pm$3. For photon $E_T>3$ GeV, $|\eta|<2.37$, $M_{\gamma \gamma}>$6 GeV, $p_T^{\gamma
\gamma}<2$ GeV:, ATLAS measured $\sigma=78 \pm 13 (stat) \pm 8 (syst)$ nb in agreement with SM (49 $\pm$ 5 nb)~\cite{atlasgamma}.

\begin{figure}[h]
\centering
\includegraphics[width=2.5in,clip]{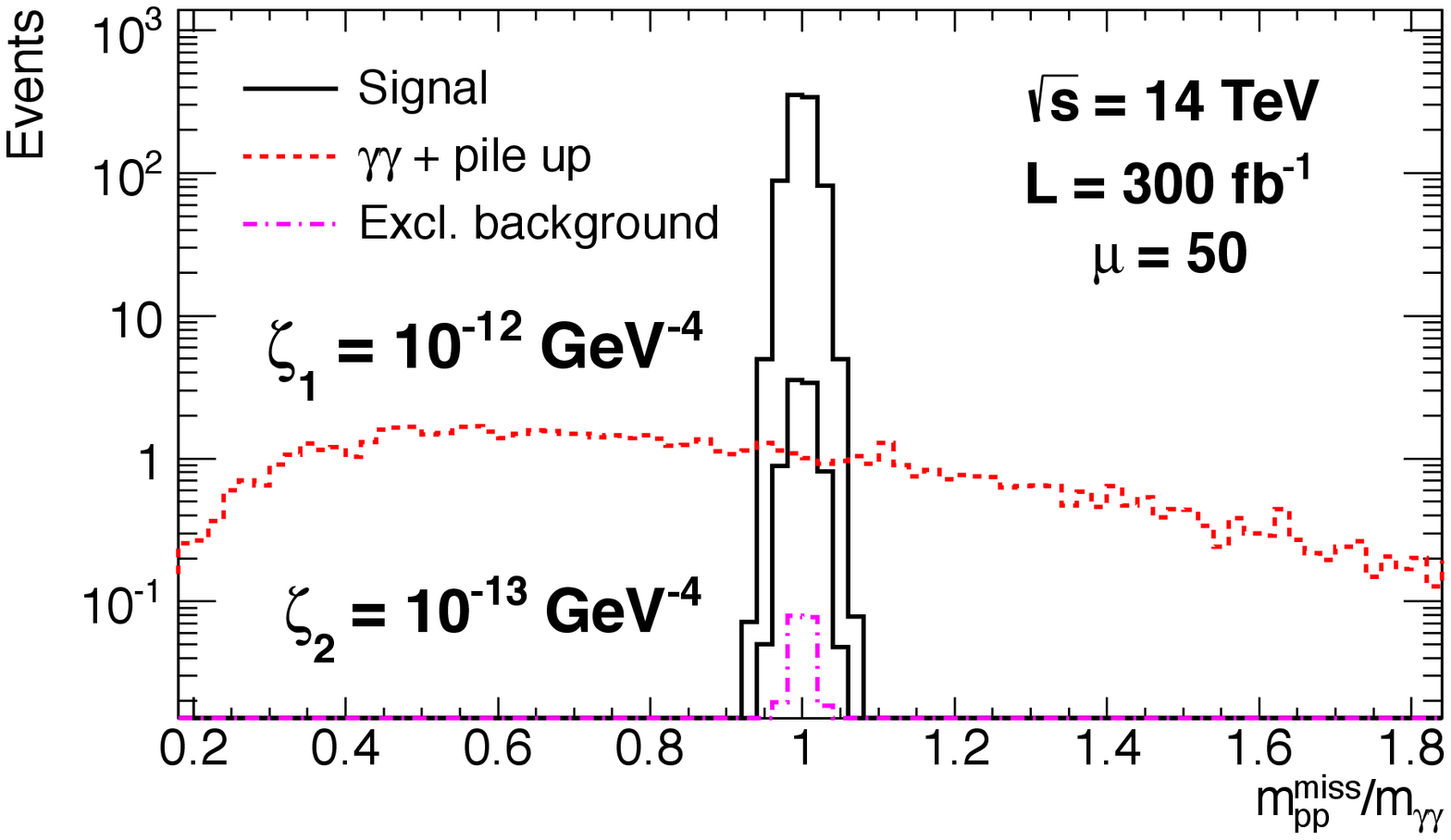}
\includegraphics[width=2.5in,clip]{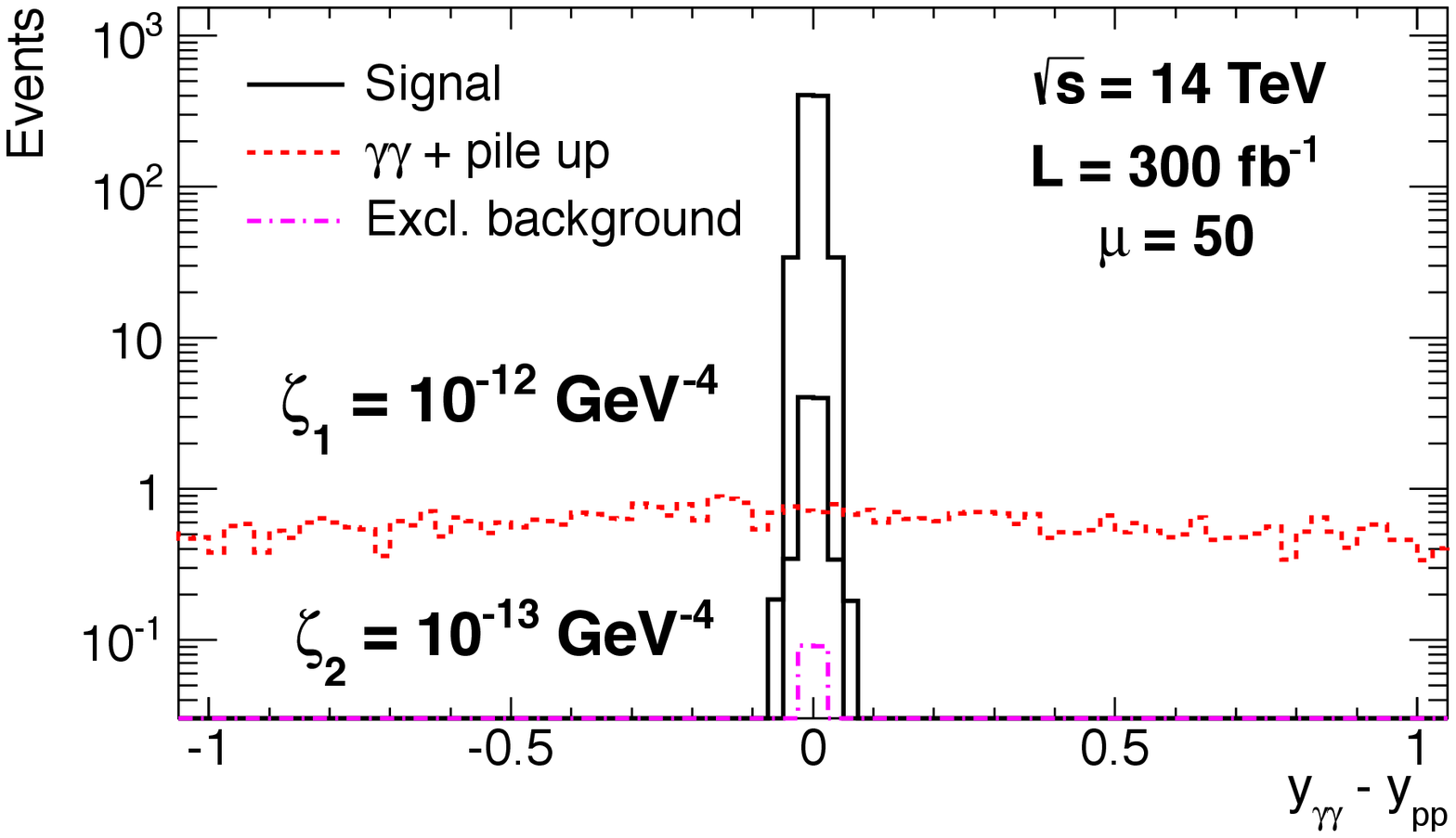}
\caption{Left: Ratio between the proton missing mass and di-photon mass for exclusive di-photon signal events and background. Right: Difference between the di-photon and di-proton rapidity for exclusive di-photon signal and background.}
\label{fig11}       
\end{figure}

Using PPS and AFP, it is also possible to search for $\gamma \gamma \gamma \gamma$ anomalous couplings in a very clean way.
Within the acceptance of the PPS and AFP detectors during standard high luminosity runs at the LHC (basically for a di-photon mass above 350 GeV), it is possible to show that the exclusive production of di-photons is completely dominated by photon exchange processes and gluon exchanges can be neglected. Since the signal only shows two photons and two intact protons in the final state, we measure all final state particles. That allows us to obtain a negligible background for 300 fb$^{-1}$ at the LHC. The basic idea is to compare the proton missing mass and the di-photon mass as shown in Fig.~\ref{fig11}, left~\cite{anomalousgamma}. The signal peaks around 1.0 and the gaussian width is due to the detector resolution whereas the pile-up background leads to a much flatter distribution since the two protons are not related with the two photons. The same requirement can be performed using the difference in rapidity between the di-photon and di-proton systems, as shown in Fig.~\ref{fig11}, right. This leads to a sensitivity better by two orders of magnitude to $\gamma \gamma \gamma \gamma$ anomalous couplings compared to other methods at the LHC without detecting intact protons~\cite{anomalousgamma}.  It is possible to show that the gain of two orders of magnitude is also valid for $\gamma \gamma WW$ and $\gamma \gamma ZZ$ anomalous couplings whereas the gain reaches three orders of magnitude for $\gamma \gamma \gamma Z$ anomalous couplings~\cite{oldaww,anomalousother}.  The search for anomalous couplings with tagged protons is now being pursued by both CMS-TOTEM and ATLAS collaborations.

\begin{figure}[h]
\centering
\includegraphics[width=3.in,clip]{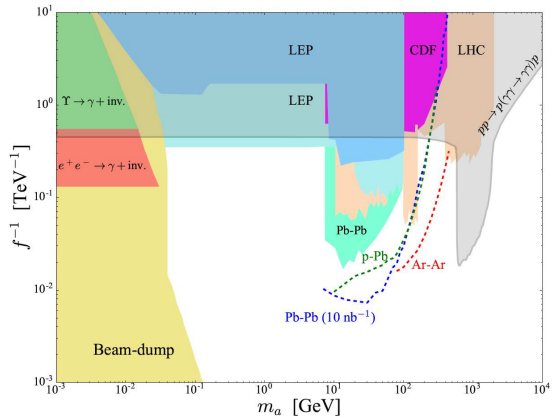}
\caption{Exclusion plot on axion like particles in the coupling versus mass plane and sensitivity at the LHC in $pp$ collisions with 300 fb$^{-1}$ (grey band) and in $PbPb$ 
(blue dashed line), $pPb$ (green dashed line), $ArAr$ (red dashed lines) collisions.}
\label{axion}       
\end{figure}

Looking for exclusive di-photon events with tagged protons can be directly applied to the search for axion-like particles at high mass~\cite{usaxion}. The ALP would be 
produced by $\gamma \gamma$ interactions and would decay into two photons. The sensitivity in the coupling versus ALP mass is shown in Fig.~\ref{axion} and we see the gain of about two orders of magnitude in coupling at high mass using this method.
We also note that this is complementary to looking for exclusive di-photons in $pPb$, $PbPb$, and $ArAr$ collisions at lower masses of ALPs. This is due to the fact that the cross section is enhanced by a factor $Z^4$ in heavy ion collisions but the sensitivity at high mass is reduced to a large suppression at small impact parameter due to the size of the heavy ion~\cite{usaxionhin}

\section{Conclusion}
We discussed many complementary measurements concerning exclusive production at the LHC from the ATLAS, CMS, LHCb and TOTEM experiments either using the ``rapidity gap" technique or the proton tags. The LHCb collaboration measured the $J/\Psi$ and $\Psi(2S)$ production and the preferred model is JMRT at NLO. The CMS exclusive pion measurement is in disagreement with theoretical expectations which is probably due to the fact that proton dissociation is not included in models. The exclusive production of $W$ pair production was observed and led to the best limits to date on $\gamma \gamma WW$ anomalous couplings. For the first time, the CMS and TOTEM collaborations observed the exclusive production of di-leptons at high mass with a proton tag. Last but not least, AFP and PPS open a completely new search for quartic anomalous couplings between $\gamma$, $W$ and $Z$ bosons with a gain in reach by two or three orders of magnitude compared to more standard methods at the LHC, reaching some of the values predicted by extra-dimension, composite Higgs models and enhancing the sensitivity to axion-like-particle at high masses.

\end{document}